\documentclass[10pt, conference, letterpaper]{IEEEtran}
\IEEEoverridecommandlockouts
    \usepackage[table]{xcolor}
    \usepackage{multirow}
\usepackage{orcidlink}
\usepackage{booktabs}
\usepackage{cite}
\usepackage{amsmath,amssymb,amsfonts}
\usepackage[nolist,nohyperlinks]{acronym}
\usepackage[inline]{enumitem}
\usepackage[detect-weight=true, detect-family=true]{siunitx}
\usepackage{algpseudocode}
\usepackage{algorithm}
\usepackage{svg}
\usepackage{multirow}
\usepackage{balance}
\usepackage{graphicx}
\usepackage[nameinlink, capitalise]{cleveref}
\usepackage[usestackEOL]{stackengine}
\DeclareSIUnit\ohm{\ensuremath\Omega}
\DeclareSIUnit\db{dB}
\DeclareSIUnit{\belmilliwatt}{Bm}
\DeclareSIUnit{\bel}{B}
\DeclareSIUnit{\bitpersecond}{bps}
\DeclareSIUnit{\samplepersecond}{Sps}
\DeclareSIUnit{\nothing}{\relax}

\usepackage{pifont}
\makeatletter
\newcommand*{\org@overidelabel}{}
\let\org@overridelabel\@verridelabel
\@ifpackagelater{acronym}{2015/03/21}{
  \renewcommand*{\@verridelabel}[1]{%
    \@bsphack
    \protected@write\@auxout{}{\string\AC@undonewlabel{#1@cref}}%
    \org@overridelabel{#1}%
    \@esphack
  }%
}{
  \renewcommand*{\@verridelabel}[1]{%
    \@bsphack
    \protected@write\@auxout{}{\string\undonewlabel{#1@cref}}%
    \org@overridelabel{#1}%
    \@esphack
  }%
}

\newcommand{\linebreakand}{%
  \end{@IEEEauthorhalign}
  \hfill\mbox{}\par
  \mbox{}\hfill\begin{@IEEEauthorhalign}
}
\makeatother
\begin{acronym}
    \acro{A-TDOA}{asynchronous \ac{TDOA}}
    \acro{ADS-TWR}{alternative \acl{DS-TWR}}
    \acro{AP-TWR}{active-passive \ac{TWR}}
    \acro{AoA}{angle of arrival}
    \acro{BLE}{Bluetooth Low Energy}
    \acro{CC-SS-TWR}{\acl{CFO} corrected \acl{SS-TWR}}
    \acro{CFO}{carrier frequency offset}
    \acro{CIR}{channel-impulse response}
    \acro{DL-TDOA}{downlink \acl{TDOA}}
    \acro{DS-TWR}{double-sided \ac{TWR}}
    \acro{GPIO}{general purpose input-output}
    \acro{IoT}{Internet of Things}
    \acro{LNA}{low-noise amplifier}
    \acro{MCU}{microcontroller unit}
    \acro{OOK}{on-off keying}
    \acro{PCB}{printed circuit board}
    \acro{PLL}{phase-locked loop}
    \acro{RSSI}{received signal strength indicator}
    \acro{RTLS}{real-time locating system}
    \acro{RTT}{round-trip time}
    \acro{SPI}{serial peripheral interface}
    \acro{SS-TWR}{single-sided \acs{TWR}}
    \acro{TDMA}{time-division multiple-access}
    \acro{TDOA}{time difference of arrival}
    \acro{TOF}{time-of-flight}
    \acro{TWR}{two-way ranging}
    \acro{USB}{universal serial bus}
    \acro{UWB}{ultra-wideband}
    \acro{WuC}{wake-up call}
    \acro{WuR}{wake-up radio}
    \acro{UAA}{uniform angular array}
    \acro{ESA}{European Space Agency}
\end{acronym}

\usepackage{eso-pic}
\usepackage{url}
\AddToShipoutPictureBG*{
  \AtPageUpperLeft{%
    \put(0,-30){\raisebox{15pt}{\makebox[\paperwidth]{\begin{minipage}{21cm}\centering
      \textcolor{gray}{This work has been accepted for presentation and publication at the 2025 IEEE International Conference on Computer Communications Workshops (INFOCOM WKSHPS), specifically the NetRobiCS 2025 workshop.} 
    \end{minipage}}}}%
  }
  \AtPageLowerLeft{%
    \raisebox{25pt}{\makebox[\paperwidth]{\begin{minipage}{21cm}\centering
      \textcolor{gray}{ \copyright 2025 IEEE.  Personal use of this material is permitted.  Permission from IEEE must be obtained for all other uses, in any current or future media, including reprinting/republishing this material for advertising or promotional purposes, creating new collective works, for resale or redistribution to servers or lists, or reuse of any copyrighted component of this work in other works.
      }
    \end{minipage}}}%
  }
}

\begin{document}
\title{WakeLoc: An Ultra-Low Power, Accurate and Scalable On-Demand \acs{RTLS} using Wake-Up Radios
\thanks{This work was funded by the \acf{ESA} under the project “Energy-aware Infrastructure-free Localization for Robotic Swarms on Lunar South Pole Missions” (OSIP Idea Id: I-2023-02020).
}
}

\author{\IEEEauthorblockN{Silvano Cortesi\orcidlink{0000-0002-2642-0797}\IEEEauthorrefmark{1}, Christian Vogt\orcidlink{0000-0003-4551-4876}\IEEEauthorrefmark{1}, and  Michele Magno\orcidlink{0000-0003-0368-8923}\IEEEauthorrefmark{1}}
\IEEEauthorblockA{\IEEEauthorrefmark{1}\textit{Department of Information Technology and Electrical Engineering, ETH Zurich, Zurich, Switzerland}} \IEEEauthorblockA{name.surname@pbl.ee.ethz.ch}
}

\maketitle
\begin{abstract}
For future large scale robotic moon missions, the availability of infrastructure-less, cheap and low power \acp{RTLS} is critical.
Traditional \ac{RTLS} face significant trade-offs between power consumption and localization latency, often requiring anchors to be connected to the power grid or sacrificing speed for energy efficiency. This paper proposes \textsc{WakeLoc}, an on-demand \ac{RTLS} based on \ac{UWB}, enabling both low-latency and ultra-low power consumption by leveraging \ac{UWB} \acp{WuR}. In \textsc{WakeLoc}, tags independently start a localization procedure by sending a \ac{WuC} to anchors, before performing the actual localization. Distributed tags equipped with \acp{WuR} listen to the \ac{WuC} and use passive listening of the \ac{UWB} messages to determine their own position. Experimental measurements demonstrate that the localization accuracy in a 2D setup achieves less than \qty{12.9}{\centi\meter} error, both for the active and the passive tag. Additional power simulations based on real-world measurements were performed in a realistic environment, showing that anchors can achieve a power consumption as low as \qty{15.53}{\micro\watt} while the \ac{RTLS} performs one on-demand localization per minute for 5 tags, thus operate up to 5.01 years on a single coin cell battery (\qty{690}{\milli\watt{}\hour}).
\end{abstract}
\acresetall

\section{Introduction}\label{sec:introduction}
The moon poses the next great challenge for both the governmental and commercial space sector, with missions such as PROSPECT by \ac{ESA} seeking to establish a lunar base for deep space missions~\cite{grenier22_posit_veloc_perfor_level_lunar, oconnor22_auton_rover_water_extrac_lunar_poles}. The lunar south pole is of specific interest due to the expected existence of ice deposits~\cite{bussey99_illum_condit_at_lunar_south_pole}. Autonomous robot swarms can pose a useful tool for discovery of such resources.

However, unlike traditional systems, these swarms must operate in infrastructure-less environments, making seamless coordination of their mission-specific subsystems, such as communication, localization, sensing, and control,  essential\cite{staudinger21_role_time_robot_swarm}. In particular, to cooperate efficiently and navigate the moon's harsh terrain, they need to be context-aware\cite{anjum24_contex}, and precise localization is a key element thereby\cite{anjum24_contex, arafat19_local_clust_based_swarm_intel}. The lack of global positioning systems or even fixed reference points on the moon amplifies this challenge and requires innovative localization strategies for precision and reliability~\cite{dabrowski08_multi_rover_navig_lunar_surfac}, while keeping the energy consumption as low as possible.

Such a space-based \ac{RTLS} must fulfill the following requirements\cite{zhang21_assem_swarm_navig_system, rocamora23_multi_robot_cooper_lunar_in}:
\begin{enumerate*}[label=\textit{\roman*)}]
\item infrastructure-less deployment: This requires ultra-low power consumption on both sides (anchors and tags) in order to achieve a fully battery-operated and easy-to-deploy system;
\item scalability on both sides (number of tags and anchors): The \ac{RTLS} should be able to handle a large number of tags, but must also work in large-scale deployments with many anchors. Thus, it needs to be collision-free and have a low coordination complexity;
\item low latency: Tags should be able to localize whenever they wish, requiring low latency or even on-demand capabilities.
\end{enumerate*}

With no access to global positioning systems in space, \ac{UWB} emerges as a compelling solution to solve these \ac{RTLS} challenges. \ac{UWB} sytems are capable of offering centimeter-level accuracy~\cite{liu24_perfor_compar_uwb_ieee} and low energy consumption in the milli-Joule range~\cite{imfeld23_evaluat_non_coher_ultra_wideb, mayer24_self_sustain_ultraw_posit_system}. However, scalability of this technology in both, the number of anchors and tags, as well as the influence of different localization schemes on power consumption is not fully researched.

In this paper we present \textsc{WakeLoc}, a \ac{WuR}-based \ac{RTLS}, combining \acp{WuR} and \ac{UWB} to a scalable and ultra-low power \ac{RTLS}. 
\acp{WuR} are ultra-low power receivers, which can wake up a host system when a specific wake-up pattern is received. These receivers operate at a fraction of the power required by traditional radio receivers, making them highly efficient for energy-constrained applications. This allows \textsc{WakeLoc} to perform ultra-low power and coordination-free localization, by combining on-demand \ac{TWR} without compromising latency, and \ac{TDOA} for large-scale scalability in the number of tags and anchors.

The advancements by \textsc{WakeLoc} can be summarized as follows:
\begin{itemize}
\item Introduction of a flexible and scalable localization scheme that allows for power reduction when using \ac{WuR}.
\item Full characterization of the system both in the real world and in simulation.
\item Comparison with the state-of-the-art \ac{RTLS} implementations \textsc{FlexTDOA} and \ac{AP-TWR} in terms of power consumption, localization accuracy and localization latency.
\end{itemize}
\section{Related Works} \label{sec:related_works}
\subsection{\acs{UWB} Localization}
\begin{table}[htpb!]
    \begin{center}
    \caption{Comparison of state-of-the-art \ac{UWB} localization schemes.}\label{tab:rel-works}
    \renewcommand{\arraystretch}{1.3}
\resizebox{\columnwidth}{!}{\begin{tabular}{@{}lrrrr@{}}
	\toprule
	\multicolumn{2}{r}{\textsc{FlexTDOA}\cite{patru23_flext}} &  \acs{AP-TWR}\cite{laadung22_novel_activ_passiv_two_way}& \multicolumn{2}{c}{\textbf{This work}}\\
	\cmidrule{4-5}
	            &                                           & \textit{active}                       & \textit{active} tag                         & \textit{passive} tag                                            \\
	\midrule
	Coord. need & Anchors                                   & Anch. + Tags                          & \cellcolor{gray!25}No                       & \multicolumn{1}{r}{\cellcolor{gray!25}No}                       \\
	On-Demand   & No                                        & \cellcolor{gray!25}Yes                & \multicolumn{1}{r}{\cellcolor{gray!25}Yes}  & \multicolumn{1}{r}{--}                                          \\
	Scalability & Anch. + Tags                              & Anch.                                 & \cellcolor{gray!25}Anch.                    & \multicolumn{1}{r}{\cellcolor{gray!25}Tags}                     \\
	Tag msg.    & \cellcolor{gray!25}\(N\) RX               & \(N\) RX, \(2\) TX                    & \(N\) RX, \(2\) TX                          & \((N+1)\) RX                                                    \\
	Anch. msg.  & \cellcolor{gray!25}\(1\) RX, \(1\) TX     & \cellcolor{gray!25}\(1\) RX, \(1\) TX & \cellcolor{gray!25}\(1\) RX, \(1\) TX       & \multicolumn{1}{r}{\cellcolor{gray!25}--}                       \\
	Acc. 2D     & --\(^a\)                                  & --\(^a\)                              & \cellcolor{gray!25}\qty{7.7}{\centi\meter}, & \multicolumn{1}{r}{\cellcolor{gray!25}\qty{9.1}{\centi\meter},} \\
	Acc. 3D     & \cellcolor{gray!25}\qty{17}{\centi\meter} & --\(^a\)                              & \(\qty{52.7}{\centi\meter}^b\)              & \(\qty{83.5}{\centi\meter}^b\)                                  \\
	\bottomrule
\end{tabular}}
    \end{center}
    \hspace{0cm}\footnotesize{\(^a\)No mean error reported (bias compensated), \(^b\)non-optimal \(z\) placement}
    \vspace{-0.2cm}
\end{table}

An overview of \ac{UWB} localization is presented in this section and~\cref{tab:rel-works}.

In \ac{TWR}, at least two messages are exchanged in order to estimate the distance between two devices using \ac{RTT} measurements. With \ac{CC-SS-TWR}, Dotlic et al. shows that it is possible to compensate for clock offsets with only two exchanged messages. In experimental tests with the \textsc{Qorvo DW1000}, the ranging error was within \qty{15}{\centi\meter} at \qty{6}{\meter} distance.

The paper of Dotlic et al. is fundamental for novel approaches on \acp{RTLS}. One of those is~\cite{laadung22_novel_activ_passiv_two_way}, where Laadung et al. present \ac{AP-TWR} methods, where either the tag actively starts ranging, or one of the anchors passively listens to the exchange, including a final report sent by the tag \(T\), to calculate the distances. They report a standard deviation below \qty{3.1}{\centi\meter} for the \textit{active} ranging tag and \qty{6.2}{\centi\meter} for the five \textit{passive} anchors. An average error is not reported. \ac{AP-TWR} approach does not scale well with the number of tags due to collisions between individual tags, and requires the anchors to be always on.

In~\cite{patru23_flext}\label{subsec:flextdoa}, Patru et al. present \textsc{FlexTDOA}, a scalable \ac{DL-TDOA}. In contrast to classic \ac{DL-TDOA}, however, the \ac{CFO} is used. Using a \ac{TDMA} schedule, a master anchor initiates the \ac{TDOA} localization. Anchors in vicinity respond to this message after a wait period \(\Delta T_i\). A \textit{passive} tag receiving all these messages can then calculate its \ac{TDOA} position using the known anchor positions \(\vec{X}_i\), and the \ac{CFO}. They achieved an accuracy of \qty{9}{\centi\meter} with a standard deviation of \qty{5}{\centi\meter} in 3D-localization using 5 anchors. This approach allows massive scalability in number of tags as well as anchors. However, the \ac{TDMA} schedule needs distribution across all anchors in order to perform localizations, increasing coordination complexity on the anchors.

To enable a scalable, coordination-less and ultra-low power localization scheme, we have combined \ac{AP-TWR} and \textsc{FlexTDOA} in \textsc{WakeLoc}, achieving scalability in both the number of tags, and anchors. By introducing our own combination of \textit{active} and \textit{passive} tags, on-demand localization capability is given with constant latency.

\subsection{Wake-up Radio Receiver}
This section presents a review of recent \acfp{WuR}, highlighting their capabilities in sensitivity, power consumption and wake-up range. An overview can be found in~\cref{tab:wur}.

Spenza et al.~\cite{spenza15_beyon} presented a \ac{WuR} for \qty{868}{\mega\hertz}, built using off-the-shelf components: A diode based envelope detector, a comparator and  an \ac{MCU} handling the signal decoding. The \ac{WuR} achieved a sensitivity of \qty{-55}{\deci\belmilliwatt} while consuming only \qty{1.276}{\micro\watt}. With a transmission power of \qty{10}{\deci\belmilliwatt} they achieved a maximal wake-up range of \qty{45}{\meter}.

In~\cite{villani24_ultra_wideb_wake_up_receiv}, Villani et al. present a \ac{WuR} for \ac{UWB} implemented as an ASIC. The chip is compatible with IEEE 802.15.4-2011 and achieves a maximum sensitivity of \qty{-86}{\deci\belmilliwatt} with a wake-up latency of \qty{524}{\milli\second} or \qty{-73}{\deci\belmilliwatt} with a latency of \qty{55}{\milli\second}. The power consumption is \qty{36}{\nano\watt}, respectively \qty{93}{\nano\watt}. The wake-up protocol is the same as in~\cite{polonelli21_ultra_low_power_wake_up}. The wake-up range is not available, as no in-field evaluations were made of the chip.

A last notable mention is \textsc{RFicient\textsuperscript{\textcopyright}}~\cite{frauenhofer_rficient}. The tri-band \ac{WuR} is capable of receiving \acp{WuC} on \qty{433}{\mega\hertz}, \qty{868}{\mega\hertz} and \qty{2.4}{\giga\hertz}. The chip achieves a sensitivity of \qty{-80}{\deci\belmilliwatt} and allows for a configurable wake-up latency from \qty{1}{\milli\second} to \qty{121}{\milli\second} leading to a power consumption between \qty{2.7}{\micro\watt} and \qty{87.3}{\micro\watt}.

\begin{table}[htpb!]
    \begin{center}
    \caption{Overview of state-of-the-art \aclp{WuR}.}
    \label{tab:wur}
    \renewcommand{\arraystretch}{1.2}
    \begin{tabular}{@{}lrrr@{}}
    \toprule
         \multicolumn{2}{r}{INFOCOM '15~\cite{spenza15_beyon}} & ISCAS '24~\cite{villani24_ultra_wideb_wake_up_receiv} & \textsc{RFicient}~\cite{frauenhofer_rficient}\\
      \midrule
      Technology & \qty{868}{\mega\hertz} & \cellcolor{gray!25}\ac{UWB} & tri-band\(^a\)\\
      Idle power & \qty{1.28}{\micro\watt} & \cellcolor{gray!25}36 -- \qty{93.2}{\nano\watt} & 2.7 -- \qty{87.3}{\micro\watt}\\
      Wake-up latency & \qty{16}{\milli\second} & 55 -- \qty{524}{\milli\second} & \multicolumn{1}{r}{\cellcolor{gray!25}1 -- \qty{121}{\milli\second}}\\
      RX sensitivity & -\qty{55}{\deci\belmilliwatt} & \cellcolor{gray!25}-86 -- -\qty{73}{\deci\belmilliwatt} & -\qty{80}{\deci\belmilliwatt}\\
      Wake-up range & \qty{45}{\meter} & - & \multicolumn{1}{r}{\cellcolor{gray!25}\qty{100}{\meter}} \\
    \bottomrule
    \end{tabular}
    \end{center}
    \hspace{0.5cm}\footnotesize{\(^a\)\qty{433}{\mega\hertz}, \qty{868}{\mega\hertz}, and \qty{2.4}{\giga\hertz}}
    \vspace{-0.2cm}
\end{table}

The presented \aclp{WuR} are subject to a trade-off between energy consumption at the transmitting node and receiving one. With respect to the initiating node's energy consumption, it would be more favorable to use a lower-frequency and narrowband \ac{WuR}. However, our simulations are based on~\cite{villani24_ultra_wideb_wake_up_receiv}, as it allows for a cheap single-transmitter solution and represents a worst-case scenario for power consumption of the initiator. It is therefore usuable as upper-bound for the efficiency of \textsc{WakeLoc}.
\section{WakeLoc}\label{sec:system-architecture}
In \textsc{WakeLoc} we combine wake-up radios on all transceivers with the advantages of the two localization concepts outlined in~\cref{sec:related_works}. These wake-up radios enable \textit{passive} listening and selective response to a request without compromising power consumption of an individual node and therefore enabling a truly event-based \ac{RTLS}, which fulfills the overall goals described in \cref{sec:introduction}.

\begin{figure}[htpb!]
    \centering
    \includesvg[width=\columnwidth]{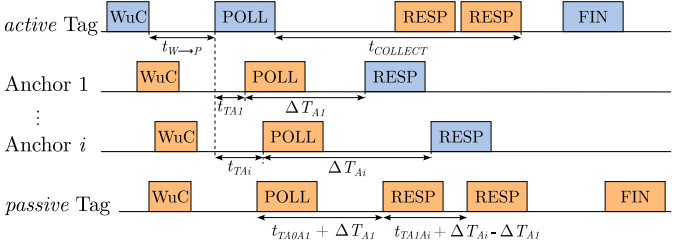}
    \vspace{-0.5cm}
    \caption{\textsc{WakeLoc} localization scheme, combining both \textit{active} \ac{AP-TWR} and \textsc{FlexTDOA}}\label{fig:new-scheme}
    \vspace{-0.3cm}
\end{figure}

\textsc{WakeLoc} tags can choose to localize themselves with two modes:  either by starting a new localization procedure on their own (\textit{active}-mode) or opportunistically waiting for another tag to start a localization procedure and make use of these messages for themselves (\textit{passive}-mode). A graphical overview of the scheme is given in~\cref{fig:new-scheme}.

\subsection{Active-mode}\label{subsubsec:active-mode}
Prior to the localization, a tag uses its radio to send a \ac{WuC} to wake up anchors (and possibly also other tags) in proximity. A tag in \textit{active}-mode (\textit{active} tag) then carries out the \textit{active} ranging from \ac{AP-TWR}~\cite{laadung22_novel_activ_passiv_two_way}. Localization is performed in a nested form of \ac{CC-SS-TWR}: In comparison to naive \ac{CC-SS-TWR}, the \textit{nested} \ac{CC-SS-TWR} reduces the distance measurement message exchange from at least \(N\) poll and \(N\) response messages to \(N+1\) messages for a localization with \(N\) anchors: all the anchors respond almost simultaneously to the same poll message, as shown in~\cref{fig:new-scheme}. This means that only \(N+1\) messages need to be exchanged in total.

After the tag has collected all distance responses, the anchors can immediately switch back to sleep and the tag calculates its own position using trilateration.

Finally, the tag sends a broadcast message containing its own estimated location -- the enabling factor for tags in \textit{passive}-mode.

\subsection{Passive-mode}
A tag in \textit{passive}-mode (\textit{passive} tag) can be in sleep mode, waiting for a \ac{WuC} from any other \textit{active} tag wanting to localize itself. After receiving the \ac{WuC}, the \textit{passive} tag listens for the poll message and the responses of all anchors in the vicinity to exploit them for \ac{TDOA}, performing the same \ac{CFO}-corrected \ac{DL-TDOA} as in \textsc{FlexTDOA} (\cref{subsec:flextdoa}) - with the difference that not an anchor is initiating the procedure, but an \textit{active} tag.

The only thing missing for a localization of the \textit{passive} tag is now the position of the \textit{active} tag. The \textit{passive} tag receives this missing information from the final message transmitted by the \textit{active} tag as mentioned in~\cref{subsubsec:active-mode}.

This scheme enables asynchronous and fully event-based \acp{RTLS}. By using \ac{WuR}, the anchors in particular can benefit from very low power consumption and localizations or messages are only exchanged when they are strictly necessary. With the addition of the \textit{passive}-mode, the \textit{active} part is being improved to such an extent that the \ac{RTLS} is scalable in terms of both the number of tags and anchors.

Despite the fact that the final message of an \textit{active} tag contains its position, its privacy can be preserved by sending its messages with a random MAC address as the source address. Then, no conclusions can be drawn about its identity.
\section{Performance evaluation setup}\label{sec:exp-setup}
The performance of the proposed \textsc{WakeLoc} system is evaluated in a scaled real-world implementation and in simulated scenarios, with the results compared to the state-of-the-art schemes described in~\cref{sec:related_works}.

\subsection{Node Hardware and Firmware}
The anchor and tag nodes consist of the same hardware, a \textsc{Qorvo DWS3000} \ac{UWB} transceiver together with an \textsc{STMicroelectronics Nucleo-L476RG} microcontroller. As the promising \ac{WuR} of Villani et al.~\cite{villani24_ultra_wideb_wake_up_receiv} is not commercially available yet, the system was emulated using \acp{WuC} generated over wires via \ac{GPIO} interrupts.  


The firmware for the nodes ensures that the \ac{MCU} is always in 'Stop 2' sleep-state when waiting for interrupts, or actively working otherwise.

The parameters of \textsc{WakeLoc} are tuned to minimize energy consumption, while still fulfilling timing requirements for the localization. The time between the transmission of the initial poll message and the anchors response is set to \(\Delta T_{A_i} = \qty{720}{\micro\second} + \qty{230}{\micro\second}\cdot i\). A wake-up time of \qty{500}{\micro\second} is assumed, once the \ac{WuC} has been sent. Using a \ac{WuC} duration of \qty{55}{\milli\second}, the latency results in \(\approx\qty{60}{\milli\second}\) for \(N=5\) anchors.

\subsection{Real-world setup}\label{subsec:characterization}
The characterization of the proposed \textsc{WakeLoc} has been performed in two steps:
\begin{enumerate*}[label=\textit{\roman*)}]
  \item characterization of the localization accuracy, and
  \item characterization of the power consumption.
  \end{enumerate*}

\begin{figure}[htpb!]
  \centering
  \includesvg[width=0.5\columnwidth]{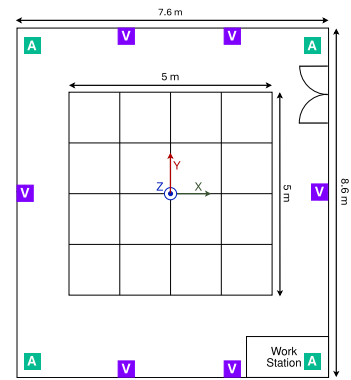}
  \caption{Experimental setup for the estimation accuracy evaluation. \(A\) marks anchors, \(V\) cameras of the \textsc{Vicon} system. The tags \(T\) were placed at different locations in the grid.}\label{fig:exp-setup}
  \vspace{-0.3cm}
\end{figure}

The real-world setup for localization accuracy is shown in~\cref{fig:exp-setup}, consisting of 4 fixed anchors and 2 moving tags. One tag is \textit{active}, one \textit{passive}. The anchors are placed in the corners of a \qtyproduct{8.6x7.6}{\meter} room and mounted between \qty{1.75}{\meter} and \qty{2.10}{\meter} above ground.
The two tags were positioned at different heights in a \qtyproduct{5x5}{\meter} grid. A total of 70 different tag positions were evaluated, and over 100 localization measurements were recorded at each position at \qty{10}{\hertz}. Ground truth positions are provided by a \textsc{Vicon} motion capture system and a \textsc{Leica Disto X310} laser distance meter.
\begin{table*}[htpb!]
  \vspace{-0.3cm}
  \renewcommand{\arraystretch}{1.2}
  \centering
  \caption{Statistics of the localization experiments. Values are given in \unit{\centi\meter}.}\label{tab:accuracy}
  \begin{tabular}{@{}lrrrrrcrrrrrcrrrrr@{}}
    \toprule
    & \multicolumn{5}{c}{\textsc{FlexTDOA}} & \phantom{a} & \multicolumn{5}{c}{\textsc{WakeLoc} \textit{active}} & \phantom{a} & \multicolumn{5}{c}{\textsc{WakeLoc} \textit{passive}} \\
    \cmidrule{2-6} \cmidrule{8-12} \cmidrule{14-18}
    & \(x\) & \(y\) & \(z\) & 2D & 3D && \(x\) & \(y\) & \(z\) & 2D & 3D && \(x\) & \(y\) & \(z\) & 2D & 3D \\
    \midrule
    avg & -9.8 & 3.0 & -32.1 & 12.9 & 43.8 && -3.0 & -1.0 & 52.7 & 7.7 & 83.5 && -5.1 & 1.3 & 24.5 & 9.1 & 82.6 \\
    md & -9.8 & 2.7 & -23.5 & 12.5 & 31.0 && -3.1 & -0.1 & -15.2 & 7.1 & 28.0 && -5.0 & 0.6 & -15.2 & 8.1 & 52.8 \\
    \(\sigma(\cdot)\) & 7.2 & 7.5 & 43.6 & 6.8 & 35.0 && 6.3 & 5.8 & 113.5 & 4.8 & 93.7 && 6.8 & 7.3 & 104.2 & 6.7 & 69.0 \\
    \bottomrule
  \end{tabular}
  \vspace{-0.2cm}
\end{table*}
The first measurements of each anchor and tag have been used for antenna delay calibration of the DW3000.

To evaluate the power consumption, the current consumption of the individual states together with their duration was recorded for both the \ac{MCU} and the \textsc{DW3000} using a \textsc{Nordic Power Profiler Kit 2}. The sampling frequency was \qty{100}{\kilo\samplepersecond}. The \ac{WuR} has already been evaluated in~\cite{villani24_ultra_wideb_wake_up_receiv}, their power consumption values have directly been used in our simulation.

\subsection{Simulation}
To determine the power consumption, the real-world scenario was simulated using Python.

\subsubsection{Node model}
To model the nodes, the power measurements described in~\cref{subsec:characterization} were accumulated along with the durations of the individual component's active states. Afterwards, the energy consumption per localization event was evaluated for each node type. The energy consumption itself is modeled depending on the number of anchors and the transmission distance of each node.

\subsubsection{Environment models}
The simulation is based on the node distribution of the \textsc{Cloves} testbed~\cite{molteni22_cloves}. The simulated environment consists of the following scenario: 
The deployment spans over two larger areas connected by a narrow region, with a size of \qty{6382}{\meter\squared} and various obstacles. Our simulation assumes sporadic localization requests, e.g. individual agents wanting to navigate in the area. In addition to the locations of the 89 anchors, the environment is evaluated for 5, 20, and 100 tags. In order for the tags to calculate their positions, a minimum amount of 5 responses per tag is assumed. 


The algorithms are compared, based on the following rationale:
\begin{description}[leftmargin=\parindent]
  \item[FlexTDOA:] The \textsc{FlexTDOA} power consumption is independent of the distribution and number of tags. It depends on the \textsc{FlexTDOA} update frequency \(T_F\) and is evaluated for  \(T_F\)= \qty{60}{\milli\second}, \qty{100}{\milli\second}, \qty{1}{\second}, \qty{10}{\second} and \qty{100}{\second}.
  \item[WakeLoc:] The anchor power consumption depends on the tag position. Therefore, 20 random tag placements were generated. For each tag placement and localization period (from \qty{60}{\milli\second} to \qty{10}{\second}), the power consumption was simulated by randomly selecting one of the tags as \textit{active} while the others are \textit{passive} - until each tag had performed 20 localizations (\textit{passive} and \textit{active} combined). This leads to an average consumption for the tags and anchors for each placement and localization period. The \ac{WuC} distance is considered to be up to \qty{20}{\meter}. 
  \item[AP-TWR:] The simulation of \textit{active} \ac{AP-TWR} is the same as the one of \textsc{WakeLoc}, except that no \ac{WuC} is transmitted and the anchors' transceivers' are always on.
\end{description}

Significantly more different localization periods were simulated for \textsc{WakeLoc} and \ac{AP-TWR} than for \textsc{FlexTDOA}. The reason for this is that \textsc{FlexTDOA} has a fixed localization period (which represents as well the latency), and in \textsc{WakeLoc} and \ac{AP-TWR} the localizations happen asynchronously - modeled as random localization periods. The minimal localization period is \qty{60}{\milli\second}, which is the shortest possible period possible when using the selected \ac{WuR}, and is thus selected for \textsc{FlexTDOA}, \ac{AP-TWR} and \textsc{WakeLoc} for comparability.
\section{Results}\label{sec:results}
\subsection{Localization Accuracy Characterization}
In our experiments' scope, a total of 9467 localizations were made across 70 different tag constellations. An overview of these results can be found in~\cref{tab:accuracy}. 
The localization results for \textit{active} \ac{AP-TWR} are the same as for \textit{active} \textsc{WakeLoc} and are thus not explicitly listed.


In the \(x\)- and \(y\)-direction, both \textsc{FlexTDOA} and \textsc{WakeLoc} (\textit{active} and \textit{passive}) show an accuracy of less than \qty{15}{\centi\meter}, which is inline with other works~\cite{ledergerber15, laadung22_novel_activ_passiv_two_way, corbalan20_ultra_wideb_concur_rangin}. With an average error of \qty{-3.0}{\centi\meter} in \(x\)-axis and \qty{-1.0}{\centi\meter} in \(y\)-axis, the \textit{active} tag in \textsc{WakeLoc} achieves the highest average 2D localization accuracy of \qty{7.7}{\centi\meter}. The standard deviation \(\sigma(\cdot)\) is \qty{4.8}{\centi\meter}. The \textit{passive} \textsc{WakeLoc} tag performs slightly worse than the \textit{active} tag with an average error of \qty{9.1}{\centi\meter} and a standard deviation of \qty{6.7}{\centi\meter} in 2D localization, as expected. The reason for this is that the error of the \textit{active} tag is propagated into the result of the \textit{passive} tag, as its estimation is used as ground-truth for the time calculations of the \textit{passive} tag. With an average error of \qty{12.9}{\centi\meter} and a standard deviation of \qty{6.8}{\centi\meter}, the \textsc{FlexTDOA} performs worst in the 2D localization, with a standard deviation similar to that of the \textsc{WakeLoc} \textit{passive} tag.

In \(z\)-axis, the \textsc{WakeLoc} \textit{active} tag has a large average error of \qty{52.7}{\centi\meter} with a standard deviation of \qty{113.5}{\centi\meter}. This results in a large error of \qty{83.5}{\centi\meter} in average with a standard deviation of \qty{93.7}{\centi\meter} for the 3D localization. This error could be improved by changing the \(z\)-placement of the anchors ~\cite{pan22_indoor_scenario_base}. The error then propagates to the \textsc{WakeLoc} \textit{passive} tag, resulting in a similar 3D localization error of \qty{82.6}{\centi\meter} in average. With the \textsc{FlexTDOA} scheme, the average error is \qty{43.8}{\centi\meter} with a standard deviation of \qty{35}{\centi\meter}.

However, it is notable that the localization of the \textsc{WakeLoc} \textit{passive} tag is on par with the \textit{active} tag: The reason for this is that the measurement noise of the \textit{active} and \textit{passive} tag are dependent random variables, since certain measurements (\(\Delta T_{A_i},\ t_{TA_i},\ \dots\)) are the same. Therefore, measurement errors might even be reduced as more data is available. This is also the reason why \textsc{WakeLoc} \textit{passive} can deliver better results than \textsc{FlexTDOA}, even though only an estimation is used for the position of the initiator instead of its exact location.

\subsection{Power characterization}
Experimental power measurements of the nodes are given in~\cref{tab:powermeasurement} and are used as basis for the power simulation in~\cref{subsec:powersim}. The \ac{WuC} by the \emph{active} Tag is with \qty{5.61}{\milli\joule} the largest contribution, compared to the approx. \qty{6}{\micro\joule} of the \emph{passive} Tag and anchor.
\begin{table*}[!htpb]
  \renewcommand{\arraystretch}{1.15}
  \begin{center}
  \vspace{-0.35cm}
  \caption{Power consumption of the \textsc{WakeLoc} protocol.}\label{tab:powermeasurement}
  \vspace{-0.08cm}
\begin{tabular}{@{}llrrr@{}}
\toprule
                              &          & \textit{active} Tag & \textit{passive} Tag & Anchor \\ 
\midrule
\multirow{2}{*}{Wake-up}      & Energy   & \qty{5.61}{\milli\joule}\(^a\) & \qty{6.6}{\micro\joule}\(^a\) & \qty{5.61}{\micro\joule}\(^a\)\\
                              & Duration & \qty{55}{\milli\second}  & \qty{55}{\milli\second}  & \qty{55}{\milli\second}\\
\midrule
\multirow{2}{*}{Localization} & Energy   & \(\qty{217.97}{\micro\joule} + N\cdot\qty{24.88}{\micro\joule}\) & \(\qty{236.97}{\micro\joule} + (N-1)\cdot\qty{24.88}{\micro\joule}\) & \(\qty{147.87}{\micro\joule} + (N-1)/2\cdot\qty{9.66}{\micro\joule}\)\\
                              & Duration & \(\qty{3.19}{\milli\second} + N\cdot\qty{210}{\micro\second}\) & \(\qty{3.15}{\milli\second} + (N-1)\cdot\qty{210}{\micro\second}\) & \(\qty{2.78}{\milli\second} + (N-1)/2\cdot\qty{230}{\micro\second}\)\\
\midrule
Sleep                         & Power    & \qty{12.05}{\micro\watt}\(^a\) & \qty{12.05}{\micro\watt}\(^a\) & \qty{12.05}{\micro\watt}\(^a\)\\
\bottomrule
\end{tabular}
\end{center}
\hspace{1.6cm}\footnotesize{\(^a\)these results contain our measurements and results of \cite{villani24_ultra_wideb_wake_up_receiv} for the \ac{WuR}}
\vspace{-0.2cm}
\end{table*}

\subsection{Power simulation}\label{subsec:powersim}
\begin{figure*}[htpb!]
  \includesvg[width=0.95\textwidth]{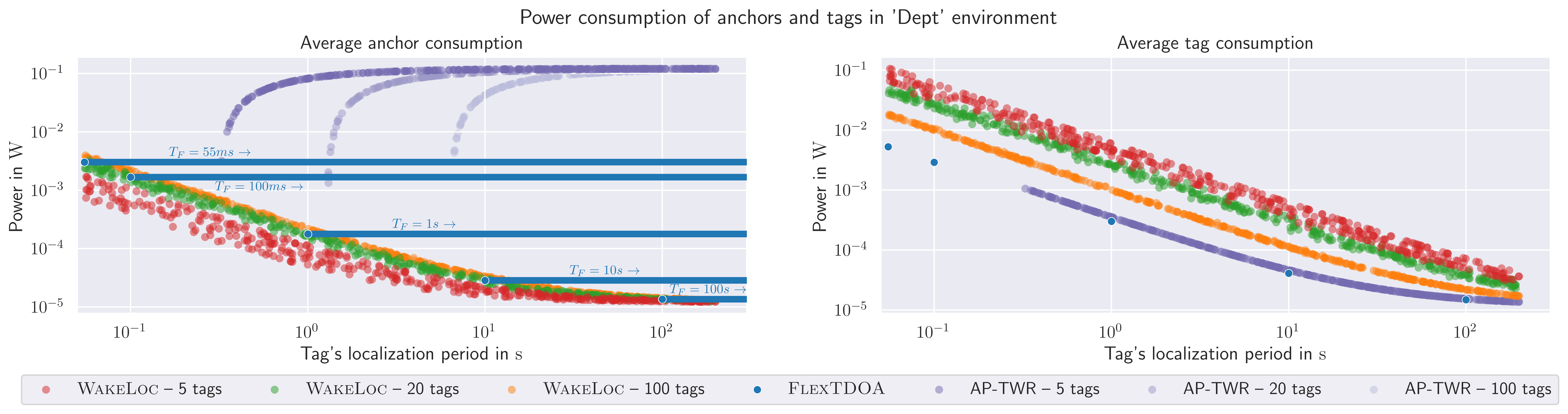}
  \vspace{-2mm}
  \caption{Power simulation. Blue line marks the consumption of \textsc{FlexTDOA} (independent on number of tags); Red, green and orange mark the consumption of \textsc{WakeLoc} in dependence of the localization period, number of tags and their positions. Fig. (a) shows the average consumption of any individual anchors. Fig. (b) shows the average consumption of an individual tag (including both \textit{active} and \textit{passive} localizations).}\label{fig:dept_power}
  \vspace{-0.5cm}
\end{figure*}

The simulated power consumption can be seen in~\cref{fig:dept_power}. 
The power consumption of the \textsc{FlexTDOA} anchors depends on the update frequency \(T_F\), but not on the tags' actual localization period, which may be lower - leading to a straight line in~\cref{fig:dept_power}(a): The line starts on the tags' maximal possible localization frequency and stays constant when tags decide to localize less often.

For \textsc{WakeLoc}, a change in the localization frequency of individual tags immediately imposes a change in the power consumption for the anchors. Additionally the placement of the tags affects both the consumption of the anchors and those of the other tags, as only anchors/tags within wake-up range are used for the localization. 


The effect of the selective anchor wake-up can be clearly seen: depending on the number of tags and their localization frequency, the power consumption of each anchor using \textsc{WakeLoc} is up to \(57.3\%\) lower than the power consumption of \textsc{FlexTDOA}: With 5 tags, \(T_F=\qty{100}{\milli\second}\) the consumption of an anchor in \textsc{FlexTDOA} is \qty{1.68}{\milli\watt}, whereas the expected consumption of an anchor in \textsc{WakeLoc} is only \qty{717}{\micro\watt}. This is a direct consequence of the large area deployment with many anchors and only few tags. With increasing number of tags, this advantage vanishes. For a worst case of 100 tags at \(T_F=\qty{1}{\second}\) the anchors have active times similar to those of \textsc{FlexTDOA}, but have an additional cost of the \ac{WuR}.

The anchors of \ac{AP-TWR} are not competitive in any scenario: Although in all cases collisions occur with low localization periods below \qty{332}{\milli\second}, the expected consumption per anchor increases logarithmically and converges to \qty{120.9}{\milli\watt}, as the idle receiver consumes more power than when packets are received and processed. 

If we assume that the localization latency should be \(T_F = \qty{1}{\second}\), and the tags requesting updates every minute, \textsc{FlexTDOA} requires significantly more energy than \textsc{WakeLoc}: For 100 tags, the \textsc{FlexTDOA} anchors consume \qty{178.5}{\micro\watt} each, \(11.5\times\) more than a \textsc{WakeLoc} anchor (\qty{15.53}{\micro\watt}). Therefore, \textsc{WakeLoc} anchors powered by a \qty{690}{\milli\watt{}\hour} CR2023 battery could operate autonomously for up to \(5.01\) years, while a \textsc{FlexTDOA} anchor would be depleted after \(161\) days.

While as few tags as possible are beneficial for the anchors, for the tags the opposite can be observed: Fewer tags increase the number of \textit{active} localizations per tag -- which is associated with high costs for the wake-up: With 5 tags, \(T_F=\qty{1}{\second}\) the power consumption is more than \(12.3\times\) higher in \textsc{WakeLoc} than \textsc{FlexTDOA} and \(10.5\times\) higher than \ac{AP-TWR}. These values make a battery-powered tag with the assumed \ac{WuR} impossible: Its runtime with the above battery would be limited to \(7.7\) hours. The more tags are available, the better the \ac{WuC} get distributed between the individual tags, leading to a lower average consumption per tag. 

To enable battery-powered tags, it is necessary to bring the tag consumption of \textsc{WakeLoc} closer to that of \textsc{FlexTDOA} (acting as a lower bound, see~\cref{tab:rel-works}). One way to achieve this would be to use \ac{WuR} that enable shorter wake-up patterns. With a decrease in wake-up duration from \qty{55}{\milli\second} to \qty{5}{\milli\second}, the advantage of \textsc{FlexTDOA} in the above example (5 tags, \(T_F=\qty{1}{\second}\)) would decrease from \(12.3\times\) to \(2.1\times\).
\section{Conclusion}\label{sec:conclusion}
In this paper we propose \textsc{WakeLoc}, an on-demand \ac{RTLS}. It makes use of \acp{WuR} and \ac{UWB}-based measurements in order to offer both -- ultra-low power consumption anchors as well as tags, while keeping the latency of the localization low.
In order to profile the novel \textsc{WakeLoc} scheme, it is compared to the state of the art \textsc{FlexTDOA} and \ac{AP-TWR}, both in experimental evaluation as well as simulations for larger scale deployments. Real-world experiments showed localization accuracies comparable to state-of-the-art systems, achieving an average accuracy below \qty{12.9}{\centi\meter} in 2D localization. In terms of average anchor power requirement, \textsc{WakeLoc} clearly outperforms \textsc{FlexTDOA} by a reduction of  \(57.3\%\) for few tags, while approximating the power consumption of \textsc{FlexTDOA} for a large number of tags. \ac{AP-TWR} is never competitive with \textsc{WakeLoc}. However, depending on the power consumption of the \ac{WuR}, the tag's power consumption is increased by \textsc{WakeLoc}. With the selective and on-demand-based system, anchors can be operated from a \qty{690}{\milli\watt{}\hour} coin-cell battery for up to 5 years, whilest the \ac{RTLS} performs on-demand localization with 5 tags and up to one localization per minute.

\section*{Acknowledgment}
The authors would like to thank Nando Galliard and Recep Polat for their commitment during their semester thesis.
\bibliographystyle{IEEEtran}
\bibliography{bib/IEEEabrv, bib/references}
\end{document}